\documentclass[10pt,twocolumn,notitlepage]{revtex4-1}
\usepackage{amsmath,amssymb,setspace}
\usepackage{graphicx}
\usepackage{newtxtext,newtxmath}
\usepackage[normalem]{ulem}
\usepackage{nicefrac}
\usepackage{mathtools}
\usepackage[utf8x]{inputenc}
\usepackage[dvipsnames]{xcolor}
\definecolor{darkblue}{rgb}{0,0,0.7}
\definecolor{darkred}{rgb}{0.5,0,0}
\usepackage[unicode,colorlinks,citecolor=darkblue,linkcolor=darkblue,bookmarks=true]{hyperref}
\newcommand{\ie}{{\it i.e.}}
\newcommand{\eg}{{\it e.g.}}
\newcommand{\etc}{{\it etc}}

\newcommand{\nf}{\nicefrac}
\newcommand{\sh}{\sinh}
\newcommand{\ch}{\cosh}

\begin{document}

\title{Gravitational wave detection beyond the standard quantum limit \\using a negative-mass spin system and virtual rigidity}
\author{Emil Zeuthen}
\email{zeuthen@nbi.ku.dk}
\affiliation{Niels Bohr Institute, University of Copenhagen, DK-2100 Copenhagen,
Denmark}
\author{Eugene S.~Polzik}
\email{polzik@nbi.ku.dk}
\affiliation{Niels Bohr Institute, University of Copenhagen, DK-2100 Copenhagen,
Denmark}
\author{Farid Ya.~Khalili}
\email{khalili@phys.msu.ru}
\affiliation{Russian Quantum Center, Moscow, Russia}

% \date{Draft of \today, \currenttime}
\date{\today}

\begin{abstract}
Gravitational wave detectors (GWDs), which have brought about a new era in astronomy, have reached such a level of maturity that further improvement necessitates quantum-noise-evading techniques. 
Numerous proposals to this end have been discussed in the literature, e.g., invoking frequency-dependent squeezing or replacing the current Michelson interferometer topology by that of the quantum speedmeter.
Recently, a proposal based on the linking of a standard interferometer to a negative-mass spin system via entangled light has offered an unintrusive and small-scale new approach to quantum noise evasion in GWDs~[Phys.\,Rev.\,Lett.\,{\bf 121}, 031101 (2018)]. The solution proposed therein does not require modifications to the highly refined core optics of the present GWD design and, when compared to previous proposals, is less prone to losses and imperfections of the interferometer. 
In the present article, we refine this scheme to an extent that the requirements on the auxiliary spin system are feasible with state-of-the-art implementations. This is accomplished by matching the effective (rather than intrinsic) susceptibilities of the interferometer and spin system using the virtual rigidity concept, which, in terms of implementation, requires only suitable choices of the various homodyne, probe, and squeezing phases.
\end{abstract}

\maketitle
% \tableofcontents

\section{Introduction}\label{sec:Intro}

The sensitivity of the contemporary state-of-art optical interferometers is to a large degree limited by quantum fluctuations of the probing light. In particular, in the modern laser interferometric gravitational wave detectors (GWDs), like Advanced LIGO~\cite{LIGOsite} and Advanced VIRGO~\cite{VIRGOsite}, the dominating noise source in the mid- and high-frequency parts of their sensitivity band (above $\sim100\,{\rm Hz}$) is the shot noise, which originates from the quantum fluctuations of the light phase~\cite{PRL_116_131103_2016}. In the more general context of the theory of linear quantum measurements~\cite{92BookBrKh, 12a1DaKh, 19a1DaKhMi}, it is known as the measurement imprecision noise. In the lower frequencies band, the technical (that is non-quantum) noise sources dominate for now. 

The resulting sensitivity has proved to be sufficient for direct observation of gravitational waves from astrophysical sources~\cite{PRL_116_131103_2016} with an event rate which exceeded one per week during the current (as of June 2019) O3 observing run of the  Advanced LIGO and Advanced VIRGO interferometers~\cite{LIGOtwitter}. At the same time, almost all GW signals detected to date came from only one class of cosmic events, namely binary black hole coalescences. In order to regularly detect gravitational waves from less powerful events, like neutron star coalescences and supernova explosions, the next major step in increasing the sensitivity of the GWDs is required.

The shot noise can be suppressed either by the brute-force increase of the optical power circulating in the interferometer or by injecting squeezed light into the interferometer dark port, as was proposed in Ref.~\cite{Caves1981}. Squeezed light is used in the smaller GW detector GEO-600 since 2011~\cite{Nature_2011, Grote_PRL_110_181101_2013}. Starting from the beginning of the O3 observing run, it is used in the  Advanced LIGO and Advanced VIRGO as well.

Due to the Heisenberg uncertainty relation, suppression of the shot noise leads to the proportional increase of another kind of quantum noise, namely the radiation pressure noise, also known as the quantum back-action noise~\cite{12a1DaKh}. It originates from the quantum fluctuations of the light power in the interferometer, which create a random force perturbing the positions of the interferometer mirrors. Within the sensitivity band of the laser GW detectors ($\gtrsim 10\,{\rm Hz}$), the suspended mirrors of the GW detectors can be treated as free masses. Correspondingly, the massless susceptibility function of the signal mechanical degree of freedom of the interferometer can be approximated as 
\begin{equation}
  \chi_I(\Omega) = -1/\Omega^2 \,,\label{eq:chi-intro}
\end{equation} 
where $\Omega$ is the observation frequency. Therefore, the radiation pressure noise is most important in the low-frequency part of the GW detectors' sensitivity band. When the Advanced LIGO reaches its design sensitivity, the radiation pressure noise will be the
dominating one at low frequencies, \ie, below $\sim100\,{\rm Hz}$.

At any given frequency $\Omega$, an optimal value of the optical power exists which provides the minimum of the sum quantum noise at this frequency. In the case where the imprecision noise and the back-action noise are uncorrelated, this minimum is known as the Standard Quantum Limit (SQL)~\cite{67a1eBr}. Being expressed in units of spectral density of the effective position noise, it is equal to~\footnote{We use the ``double-sided'' convention for the spectral densities in this paper; our main result, namely the value of the sensitivity gain $G$, evidently does not depend on this choice.}
\begin{equation}
  S_{\rm SQL} = \frac{\hbar}{m\Omega^2} \,.
\end{equation} 
Several methods of overcoming the SQL suitable for laser interferometers are known; see, \eg, the reviews \cite{12a1DaKh, 19a1DaKhMi}. In particular, as early as in 1982, W.\,Unruh~\cite{Unruh1982} had shown that, injecting into the interferometer squeezed light with the optimally tuned frequency-dependent squeeze angle (i.e., phase squeezing at higher frequencies and amplitude squeezing at lower ones), it is possible to suppress the quantum noise spectral density by the squeeze factor $e^{2r}$ over the entire band of interest. A practical method for generating frequency-dependent squeezed light was proposed by J.\,Kimble and co-workers in Ref.~\cite{02a1KiLeMaThVy}. They have shown that the necessary frequency dependence can be created by reflecting an ordinary frequency-independent squeezed vacuum from an additional Fabry-P\'{e}rot cavity (a so-called filter cavity). 

This scheme is considered now as one of the most probable candidates for implementation in the next generation of GW detectors. However, it has a significant disadvantage, namely, vulnerability to optical losses in the filter cavity. In order to ``dilute'' them, long (and therefore expensive) filter cavities with high-reflectivity mirrors have to be used. In fact, filter cavities with the same 4\,km length as the main interferometer arms were considered in Ref.~\cite{02a1KiLeMaThVy}. Currently, a more modest but still long (tens of meters) cavity is discussed as an option for the future upgrade of the Advanced LIGO  detectors~\cite{Evans_PRD_88_022002_2013, Kwee_PRD_90_062006_2014}.

In Ref.~\cite{Ma_NPhys_13_776_2017}, a different approach to the preparation of the necessary quantum state of light was proposed. In this scheme, two entangled light beams are prepared using an optical parametric oscillator. One of them (``signal'') probes the interferometer, and the second one (``idler'') the filter cavity. The output beams are then measured by two homodyne detectors. Due to the entanglement, measurement of the ``idler'' beam prepares the ``signal'' beam in the required frequency-dependent squeezed quantum state. Taking into account that the wavelengths of the signal and the idler beam could be different (the non-degenerate regime), some additional mode of the interferometer can be used as the filter cavity; it is this option that was considered in detail in Ref.~\cite{Ma_NPhys_13_776_2017}. This scheme does not require a dedicated filter cavity, but it requires instead  the additional squeezed light injection and the additional readout optical paths which could hinder its practical application.  

Instead of the passive filter cavity, a much more compact active ``negative effective mass'' atomic spin ensemble has been shown to cancel quantum backaction noise, generate entanglement, and perform sensing beyond the SQL. First experiments were performed with purely atomic systems~\cite{Julsgaard2001, Wasilewski2010}. Later, the idea was applied to a mechanical system~\cite{Polzik_AnnPhys_527_A15_2014} in the spirit of trajectories without quantum uncertainties based on the establishment of entanglement between a mechanical oscillator and a spin system~\cite{Hammerer_PRL_102_020501_2009}. There it has been shown that an ensemble of spins oriented (anti)parallel to the axis of the magnetic field behaves as an effective (positive-) negative-mass oscillator within the Holstein-Primakoff approximation. Those early papers utilized the ``sequential'' layout, where the same light interacts with the mechanical mode and the spin ensemble. Recently, suppression of the back-action noise using the atomic spin ensemble scheme was demonstrated experimentally, with a nanomechanical membrane playing the role of the mechanical object~\cite{Moeller_Nature_547_191_2017}.

This scheme cannot be used directly in the laser GW detectors, because, in order to interact effectively with the atomic spin ensemble, the optical wavelength must be close to that of the atomic transition ($\lambda_S\approx850\,{\rm nm}$), while in the contemporary GW detectors the wavelength is equal to $\lambda_I=1064\,{\rm nm}$ and longer wavelengths are planned for future upgrades. To circumvent this problem, a ``parallel'' layout (similar to the one of Ref.~\cite{Ma_NPhys_13_776_2017}) was proposed in Ref.~\cite{18a1KhPo}. It was shown that, using demanding but realistic parameters of the spin system, it is possible to improve the sensitivity by 6\,dB over the entire frequency band of the GWD.

In order to effectively suppress the quantum noise, two conditions have to be satisfied for the interferometer and the spin system. First, the readout rates (the measurement strengths) in the subsystems have to match each other. Second, the susceptibility of the spin system has to match that of the relevant mechanical degree of freedom in the interferometer. In Ref.~\cite{18a1KhPo}, a brute-force approach to satisfying these conditions was used, which resulted in a very demanding value of the quantum cooperativity factor $\mathcal{C}_S$ of the spin system (denoted as $d_0$ in Ref.~\cite{18a1KhPo}), about $10^2$ (see details in Sec.~\ref{sec:matching_conds}). Such a high value of $\mathcal{C}_S$ requires that the spin ensemble should be placed inside an optical cavity, which inevitably increases the optical losses, greatly hindering the implementation of this scheme. 

\begin{figure}[tb]
  \includegraphics[viewport=66bp 0bp 906bp 540bp,clip,width=\columnwidth]{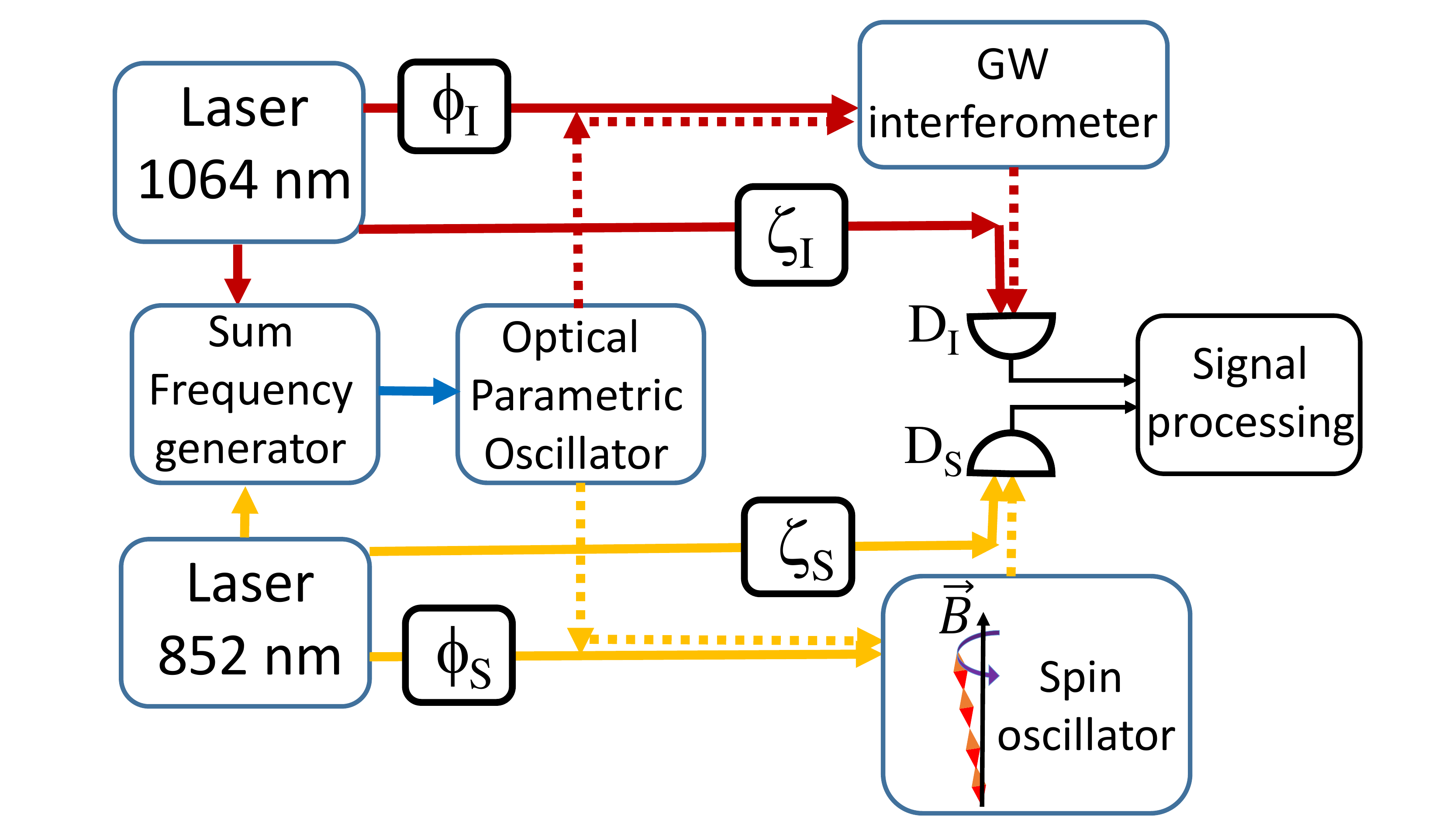}
  \caption{The GW interferometer ($I$) and the spin oscillator ($S$) are probed and detected in parallel by laser beams (solid arrows) with different wavelengths but entangled fluctuations (dashed arrows). These two-mode-squeezing correlations are achieved by means of a sum frequency generator in combination with an optical parametric oscillator. The collective spin of the atomic ensemble precesses around the magnetic field $\vec{B}$ forming a spin oscillator. With respect to the squeeze angle of this process, we reference the probe $\phi_{I,S}$ and detection $\zeta_{I,S}$ phases of the respective systems. The output fields impinge on detectors $D_{I,S}$ and the resulting measurements are suitably combined to cancel the joint meter noise. A possible practical implementation of the present conceptual schematic is presented in Ref.~\cite{18a1KhPo}.
  }\label{fig:the_scheme}
\end{figure} 

In the present paper we introduce an additional mechanism, the \emph{virtual optical rigidity} effect (see Sec.~4.4 of Ref.~\cite{12a1DaKh}), into the combined GWD and spin system. We show that this idea dramatically relaxes the requirement on $\mathcal{C}_S$. The new mechanism can be implemented by experimentally straightforward phase shifts of the optical carrier and homodyne detection. In the following we provide the general analysis of the ``parallel'' spin-system-based scheme shown in Fig.~\ref{fig:the_scheme}, assuming full freedom in our choice of the relevant phases, namely two homodyne angles and two optical carriers relative to the squeezing phase. We then show that the virtual rigidity effect can induce effective frequency shifts in the mechanical and spin system susceptibilities. Finally, we derive a simple closed equation~\eqref{Gapp} for the sensitivity gain provided by our scheme, which clearly shows the comparative role of the optical and the spin system losses at different frequencies.

The paper is organized as follows. In Sec.~\ref{sec:QN-matching-intro}, we familiarize the readers with the ``brute-force'' matching conditions of Ref.~\cite{18a1KhPo}, introduce the virtual rigidity concept and use it to derive conditions for quantum noise suppression under ideal conditions. In Sec.~\ref{sec:exact-S}, we present the full expression for the sensitivity of our scheme accounting for various imperfections, including finite entanglement between the probe fields, optical losses, the spin system thermal noise, and response mismatch due to the spin system dissipation. We then evaluate the sensitivity in Sec.~\ref{sec:plots} using state-of-the-art parameters for interferometer and spin systems to assess the potential of our scheme. Finally, we conclude and give an outlook in Sec.~\ref{sec:conclusion}.

The main parameters used in this paper are listed in Table~\ref{tab:notations}. For the GW interferometer parameters, we use the values which correspond to the Advanced LIGO design goal~\cite{CQG_32_7_074001_2015}. We would like to note, however, that our results explicitly depend only on one parameter of the interferometer, namely, the readout rate 
\begin{equation}
  \Omega_{qI} = \sqrt{\frac{16\omega_oI_c}{mcL\gamma}} \,,
\end{equation} 
which is equal to $\Omega_{qI}\approx2\pi\times60\,{\rm Hz}$ for the parameters listed in Table~\ref{tab:notations}. Taking into account that for future interferometers higher optical power but at the same time heavier test masses and longer arms are planned, it is reasonable to expect that this parameter will not change drastically. 
For the intrinsic spin linewidth we assume $\gamma_{S,0}=2\pi \times 1\,\text{Hz}$ as has been demonstrated in experiment~\cite{Balabas2010}.

\begin{table*}
  \begin{ruledtabular}
    \begin{tabular}{lll}
      Notation & Quantity & Value used for estimates\\
      \hline
      $\omega_o$ & Optical frequency & $2\pi c/(1064\,{\rm nm})$ \\
      $L$ & Interferometer arm length & 4000\,m \\
      $m$ & Mirror mass  & 40\,kg \\
      $\gamma$ & Interferometer bandwidth (half width at half maximum, HWHM) & $2\pi\times500\,{\rm Hz}$ \\
      $I_c$ & Optical power circulating in each of the arms & 840\,kW \\ 
      $\gamma_{S,0}$ & Spin system ``dark'' damping rate (HWHM) & $2\pi\times1\,{\rm Hz}$ \\
      $\eta_{iI}$, $\eta_{oI}$ & Input and output quantum efficiencies  of the interferometer & 0.95 \\
      $\eta_{iS}$, $\eta_{oS}$ & Input and output quantum efficiencies  of the spin system & 0.95 \\
    \end{tabular}
  \end{ruledtabular}
  \caption{The main parameters and their numerical values used in this paper.}\label{tab:notations}
\end{table*}

\section{Quantum noise matching}\label{sec:QN-matching-intro}

\subsection{Matching conditions for the interferometer and spin systems without virtual rigidity}\label{sec:matching_conds} 

A schematic of the experimental layout is shown in Fig.~\ref{fig:the_scheme}. To set the stage for this extended scheme, we start by reviewing the simpler scheme introduced in Ref.~\cite{18a1KhPo}. In essence, it differs from the one considered here and shown in Fig.~\ref{fig:the_scheme} only by the specific choice of the homodyne and the carrier angles: 
\begin{equation}
  \phi_I = \phi_S = 0 \,, \quad \zeta_I = \zeta_S = \frac{\pi}{2} \,.
\end{equation}
In this limiting case, relations for the input $\hat{a}_{I,S}$ and output $\hat{b}_{I,S}$ light quadratures for the interferometer and spins, respectively, are~\cite{18a1KhPo}
\begin{subequations}\label{io_prev} 
	\begin{gather}
	\hat{b}_I^{s} = 
	\hat{a}_I^{s}
	+ \chi_I\biggl(
	\frac{\Omega_{qI}}{\sqrt{\hbar m}}\,F_s + \Omega_{qI}^2\hat{a}_I^{c}
	\biggr) \,, \\
	\hat{b}_S^{s} = 
	\hat{a}_S^{s}
	+ \chi_S\bigl(\Omega_{qS}\hat{f}_T + \Omega_{qS}^2\hat{a}_S^{c}\bigr) \,, 
	\end{gather}
\end{subequations}
where $\Omega_{qS}$ and $\Omega_{qI}$ are the readout rates, $F_s$ is the signal (GW) force, $\hat{f}_T$ is the normalized thermal noise of the spin system, $\chi_I$ and $\chi_S$ are the massless susceptibilities, and the superscripts $ c $ and $ s $ denote the cosine (amplitude) and sine (phase) quadratures, respectively. 

The analysis of the work \cite{18a1KhPo} was based on two conditions for the negative-mass spin system parameters, which allow to provide the required conditional frequency-dependent squeezing across the entire bandwidth of interest. The first one is the equality of the readout rates in the two subsystems:
\begin{equation}
  \Omega_{qS} = \Omega_{qI} \,, \label{eq:theta-matching-intuitive}
\end{equation} 
where 
\begin{equation}
  \Omega_{qS} = \sqrt{4\Gamma_S\Omega_S} 
\end{equation} 
is the spin system readout rate, $\Omega_S$ is the spin system (Larmor) resonance frequency, and $\Gamma_S$ is the spin-light coupling factor proportional to the optical power probing the spin system. Both $\Omega_S$ and $\Gamma_S$ are typically highly tunable parameters for a collective spin oscillator \cite{Moeller_Nature_547_191_2017}, the former by the DC magnetic bias field and the latter by the optical probe power. However, the readout of the spin oscillator entails an increase of its bandwidth beyond its intrinsic value $\gamma_{S,0}$ due to induced spontaneous emission (i.e. power broadening)
\begin{equation}
\gamma_S-\gamma_{S,0}=\gamma_{S,\text{pb}}=\Gamma_S/\mathcal{C}_S\, , \label{eq:power-broad}
\end{equation}
where the spin oscillator cooperativity $\mathcal{C}_S$ depends on factors such as atomic density, the atomic species, optional cavity enhancement, and probe detuning from atomic resonance but is independent of probe power. Since it is desirable to keep the spin decay $\gamma_S$ small, due to the noise and response mismatch it otherwise entails, we are motivated to realize the condition~\eqref{eq:theta-matching-intuitive} by means of a relatively small $\Gamma_S$ (so as to keep $\gamma_{S,\text{pb}}$ small) and a large $\Omega_S$.
However, this strategy raises another problem pertaining to the second matching condition which must be fulfilled for broadband noise cancellation:
\begin{equation}
  \chi_I (\Omega)= \frac{1}{-\Omega^2} 
  \approx -\chi_S (\Omega) = \frac{1}{\Omega_S^2 - \Omega^2 - 2i\gamma_S \Omega} \, ,
  \label{eq:chi-matching-intuitive}
\end{equation}
where $\chi_S$ is the spin system (massless) susceptibility. In addition to showing the need for small $\gamma_S$, this condition prompts us to employ a small $\Omega_S \rightarrow 0$, contrary to what was suggested by the first condition~[Eq.~\eqref{eq:theta-matching-intuitive}]. These two opposing requirements can, in principle, be accommodated by a compromise involving a small, finite $\Omega_S$ as in the original proposal~\cite{18a1KhPo}, where the value $2\pi\times 3\,\text{Hz}$ was used.
But this strategy demands a highly refined spin system with $\mathcal{C}_S\sim10^2$ to avoid an excessively power-broadened $\gamma_S$~\eqref{eq:power-broad}, thus posing a significant practical challenge.

\subsection{Virtual rigidity representation}\label{sec:virtual-rigidity-rep}

Let us start with a standard optomechanical setup consisting of a mechanical object (free mass or harmonic oscillator) whose motion modulates the eigenfrequency of an optical cavity probed by a pump laser. Using the well-known analogy between the mechanical system and the collective spin mode of the atomic ensemble \cite{Hammerer_RMP_82_1041_2010, Moeller_Nature_547_191_2017}, our treatment here can be extended to the latter, as we will make use of in Sec.~\ref{sec:noise-cancel-intuitive}. For simplicity, we neglect here the optical and mechanical losses (they will be taken into account later). Also for simplicity, we assume that our frequency band of interest is well within the cavity half-bandwidth $\gamma$ (the bad-cavity approximation): 
\begin{equation}\label{bad_cavity} 
  \Omega \ll \gamma \,.
\end{equation} 
This assumption, which we retain throughout this work, will be discussed in more detail in Sec.~\ref{sec:exact-S-intro}.

The input-output relation for this system can be presented as follows:
\begin{equation}\label{b_zeta} 
  \hat{b}^\zeta = \hat{b}^c\cos\zeta + \hat{b}^s\sin\zeta
  = \hat{a}^\zeta
    + \chi\biggl(\frac{\Omega_q}{\sqrt{\hbar m}}F_s + \Omega_q^2\hat{a}^\phi\biggr)
        \sin(\zeta-\phi) \, ;
\end{equation} 
see, \eg, the review papers \cite{12a1DaKh, 16a1DaKh}. Here $\zeta$ is the homodyne angle, $\phi$ is the optical carrier phase, $\hat{a}^{c,s}$ are the cosine and the sine quadratures of input light, respectively, satisfying the commutation relation $[\hat{a}^c, \hat{a}^s] = i$ \cite{Caves1985}, $\hat{b}^{c,s}$ are the corresponding amplitudes of the output light, and $\hat{a}^\psi$, $\hat{b}^\psi$, \etc., are the rotated quadratures in terms of an angle $\psi$, \eg ,
\begin{equation}
  \hat{a}^\psi = \hat{a}^c\cos\psi + \hat{a}^s\sin\psi \,.
\end{equation} 
It is easy to see that Eqs.~\eqref{io_prev} correspond to the particular case of Eq.~\eqref{b_zeta} where $\phi=0$ and $\zeta=\pi/2$.

From the output field~\eqref{b_zeta} the signal force is estimated as
\begin{equation}
  \tilde{F}_s = F_s + \sqrt{\hbar m}\hat{f} \,,
\end{equation} 
where
\begin{equation}\label{eq:F-sum}
  \hat{f} = \frac{\chi^{-1}}{\Omega_q\sin(\zeta-\phi)}\,\hat{a}^\zeta 
  + \Omega_q\hat{{\rm a}}^\phi
\end{equation} 
is the normalized sum quantum noise of our system. The two components of this equation describe the imprecision noise and the back-action noise, respectively; in general, these are correlated with each other. 

The consequences of the correlations can be elucidated by introducing the orthonormal quadrature basis defined by the detection angle $\zeta$:
\begin{subequations}\label{a_cs_prime} 
  \begin{gather}
    \hat{a}^c{}' \equiv \hat{a}^{(\zeta-\pi/2)}
      = \hat{{\rm a}}^c\sin\zeta - \hat{{\rm a}}^s\cos\zeta \,,\\
    \hat{a}^s{}' \equiv \hat{a}^\zeta 
      = \hat{{\rm a}}^c\cos\zeta + \hat{{\rm a}}^s\sin\zeta \,.
  \end{gather}
\end{subequations}
The resulting form of $\hat{f}$~\eqref{eq:F-sum} is 
\begin{equation}\label{eq:F-sum-virtual}
  \hat{f} = \frac{\chi_{\text{eff}}^{-1}}{\beta}\,\hat{a}^s{}' + \beta\hat{a}^c{} '\,,
\end{equation}
where 
\begin{equation}
  \beta = \Omega_q\sin(\zeta-\phi) 
\end{equation} 
is the effective readout rate and
\begin{equation}
  \chi_{\text{eff}}^{-1} \equiv \chi^{-1} + \frac{\Omega_q^2}{2}\sin2(\zeta-\phi)
  \label{eq:chi-eff}
\end{equation}
is the {\it effective susceptibility} of the scheme.
It follows from Eq.~\eqref{eq:F-sum-virtual} that the orthonormal basis~\eqref{a_cs_prime} introduces the effective, \emph{uncorrelated} imprecision and back-action noise terms constructed by absorbing the part of the back action correlated with the nominal imprecision noise into an effective imprecision noise term. A by-product of this transformation is the modified effective susceptibility $\chi_{\rm eff}$. Since the modification term in Eq.~\eqref{eq:chi-eff} is real and independent of the Fourier frequency, it corresponds to a shift in the resonance frequency of $\chi$, whence it is referred to as \emph{virtual rigidity} (see Sec.~4.4 of Ref.~\cite{12a1DaKh}). 

We conclude that, in the absence of optical losses, the probing of a system characterized by $\chi$ and probe parameters $\Omega_q$, $\zeta$, and $\phi$ is indistinguishable from the scenario resulting from using the scheme with effective parameters $\chi_{\text{eff}}$, $\Omega_{q\text{eff}}=\beta$, and $\zeta_\text{eff}-\phi_\text{eff}=\pi/2$.

\subsection{Quantum noise matching using virtual rigidity for two systems probed by entangled light}\label{sec:noise-cancel-intuitive}

Consider now two systems---the interferometer and the spin system, denoted by the subscripts $I$ and $S$, respectively. Suppose that they are probed by individual optical meters, described by the four quadratures $\hat{a}_{I}^{c,s}$ and $\hat{a}_{S}^{c,s}$, in the manner described in Sec.~\ref{sec:virtual-rigidity-rep}. The two systems are not interacting, but the fluctuations of the two light meters are assumed to be entangled by a non-degenerate parametric down-conversion process (see Fig.~\ref{fig:the_scheme}). We will use the parametric pump phase as the phase reference point (that is, the squeeze angle defines zero phase). In this case, the noise spectral densities of all four quadratures are equal to 
\begin{subequations}\label{eq:entanglement-corr}
  \begin{gather}
    S_{{\rm a}_I^c} = S_{{\rm a}_I^s} = S_{{\rm a}_S^c} = S_{{\rm a}_S^s} 
      = \frac{\ch2r}{2} \,,\label{eq:entanglement-corr_a} \\
    \intertext{and the cosine quadratures of the beams are correlated whereas the sine counterparts are anti-correlated:}
    S_{{\rm a}_I^c{\rm a}_S^c} = -S_{{\rm a}_I^s{\rm a}_S^s} = \frac{\sh2r}{2} \,, 
      \label{cross_corr} 
  \end{gather}
\end{subequations}
where $r$ is the squeeze factor. All other components of the correlation matrix vanish. 

Assume now that we wish to measure a force signal acting on system $I$. Due to the meter noise correlations, the sensitivity to this signal can be improved by exploiting the additional probing on system $S$ by adding the corresponding additional output signal to the main interferometer signal with some optimal weight function $\Lambda$. 
The equations for the normalized meter noise \eqref{eq:F-sum} for the two systems are 
\begin{subequations}\label{f_IS} 
  \begin{gather}
    \hat{f}_I 
      = \frac{\chi_I^{-1}}{\Omega_{qI}\sin(\zeta_I-\phi_I)}\,\hat{a}_I^{\zeta_I}
        + \Omega_{qI}\hat{a}_I^{\phi_I} \,, \\
    \hat{f}_S 
      = \frac{\chi_S^{-1}}{\Omega_{qS}\sin(\zeta_S-\phi_S)}\,\hat{a}_S^{\zeta_S}
        + \Omega_{qS}\hat{a}_S^{\phi_S} \,,
  \end{gather}
\end{subequations}
where $\zeta_{I,S}$ are the homodyne angles in the interferometer and the spin system channels, respectively, and $\phi_{I,S}$ are the corresponding probe phases of the carrier fields. In Ref.~\cite{18a1KhPo}, the simplest particular case of measuring the phase quadratures of light while probing with the (orthogonal) amplitude fluctuations was considered,
\begin{equation}\label{eq:PRL-phases}
  \zeta_I = \zeta_S = \frac{\pi}{2} \,, \quad \phi_I = \phi_S = 0 \,,
\end{equation} 
which corresponds to 
\begin{subequations}
  \begin{gather}
   % \hat{f}_I = \frac{\chi_I^{-1}}{\Omega_{qI}}\,\hat{a}_I^{\zeta_I} + \Omega_{qI}\hat{a}_I^{\phi_I} \,, \\
     \hat{f}_I = \frac{\chi_I^{-1}}{\Omega_{qI}}\,\hat{a}_I^{s} + \Omega_{qI}\hat{a}_I^{c} \,, \\
%    \hat{f}_S = \frac{\chi_S^{-1}}{\Omega_{qS}}\,\hat{a}_S^{\zeta_S} + \Omega_{qS}\hat{a}_S^{\phi_S} \,.
    \hat{f}_S = \frac{\chi_S^{-1}}{\Omega_{qS}}\,\hat{a}_S^{s} + \Omega_{qS}\hat{a}_S^{c} \,.
  \end{gather}
\end{subequations}
Taking into account Eqs.~\eqref{eq:entanglement-corr}, it is easy to see see that if 
\begin{equation}\label{eq:match-cond-std}
  \Omega_{qS}^2 = \Omega_{qI}^2 \quad \text{and} \quad \chi_S = -\chi_I \,,
\end{equation} 
then the simple subtraction of $\hat{f}_S$ from $\hat{f}_I$ (which corresponds to the choice of the above-mentioned weight function $\Lambda=-1$) allows us to reduce the resulting noise spectral density by the factor $e^{2r}/2$ relative to a standard interferometer subject to vacuum noise. The optimal weight function $\Lambda=-\tanh2r$ gives the slightly better suppression factor $\cosh2r$.

As was discussed in Sec.~\ref{sec:matching_conds}, implementation of the near antisymmetric susceptibilities~\eqref{eq:match-cond-std} could be problematic due to technological limitations. However, the virtual rigidity approach can be used to make the effective susceptibilities~\eqref{eq:chi-eff} match each other. Here we demonstrate how the complete cancellation of quantum noise at all Fourier frequencies can be engineered in the limit of a lossless negative-mass spin system with
\begin{subequations}\label{chis_ideal} 
  \begin{equation}
    \chi_S^{-1} = -(\Omega_S^2-\Omega^2) \, ,
  \end{equation} 
where the overall sign stems from the negative mass (the general case  is  considered in Sec.~\ref{sec:exact-S}). For the interferometer, we suppose that
  \begin{equation}
    \chi_I^{-1} = -\Omega^2 
  \end{equation} 
  (the ideal free mass).
\end{subequations}

In general, the phase rotation transformations \eqref{a_cs_prime} in terms of $\zeta_{I,S}$ alter the cross-correlation entries in the spectral density matrix of the light quadratures. However, an interesting feature of the two-mode squeezed light generated in a non-degenerate parametric process is that if the homodyne angles $\zeta_I$ and $\zeta_S$ are antisymmetric with respect to the phase of the parametric pump (modulo $\pi$),
\begin{equation}\label{cond1_raw} 
  \zeta_I + \zeta_S = \pi n \,,
\end{equation} 
where $n$ is an integer, then the matrix remains invariant and, in particular, Eqs.~\eqref{eq:entanglement-corr} remain valid. This follows from the simple geometrical observation that Eq.~\eqref{cross_corr} implies that the quadrature pairs $(\hat{a}_{I}^{c},\hat{a}_{S}^{c})$ and $(\hat{a}_{I}^{s},-\hat{a}_{S}^{s})$ are correlated. The minus sign in the latter pair effectively inverts the sense of rotation~\eqref{a_cs_prime}, $\zeta_S\rightarrow-\zeta_S$, resulting in the antisymmetric condition~\eqref{cond1_raw} for invariance of the correlation matrix. Let us, without loss of generality, take $n=1$ in Eq.~\eqref{cond1_raw} in order to provide a smooth transition from the case of Ref.~\cite{18a1KhPo}, as stated in Eqs.~\eqref{eq:PRL-phases}.

Assuming the condition~\eqref{cond1_raw}, we can rewrite Eqs.~\eqref{f_IS} in the virtual rigidity form~\eqref{eq:F-sum-virtual}:
\begin{subequations}\label{f_IS_rotated}
  \begin{gather}
    \hat{f}_I 
      = \frac{\chi_{\text{eff}I}^{-1}}{\beta_I}\,\hat{a}_I^s{}' + \beta_I\hat{a}_I^c{}'\,,
      \label{eq:F-sum-virtual-I}\\
    \hat{f}_S 
      = \frac{\chi_{\text{eff}S}^{-1}}{\beta_S}\,\hat{a}_S^s{}' + \beta_S\hat{a}_S^c{}'\,,
      \label{eq:F-sum-virtual-S}
  \end{gather}\label{eq:F-sum-virtual-2}
\end{subequations}
where
\begin{equation}
  \beta_{I,S} = \Omega_{qI,S}\sin(\zeta_{I,S}-\phi_{I,S})
\end{equation} 
and
\begin{subequations}
  \begin{gather}
    \chi_{\text{eff}I}^{-1} = -\Omega^2 + \frac{\Omega_{qI}^2}{2}\sin2(\zeta_I-\phi_I)
      \,,\\
    \chi_{\text{eff}S}^{-1} = \Omega^2 - \Omega_S^2 
      + \frac{\Omega_{qS}^2}{2}\sin2(\zeta_S-\phi_S) \,.\label{eq:chi-eff-S}
  \end{gather}
\end{subequations}
Therefore, the setting 
\begin{equation}\label{cond2_raw} 
  \frac{\Omega_{qI}^2}{2}\sin2(\zeta_I-\phi_I) 
  = \Omega_S^2 - \frac{\Omega_{qS}^2}{2}\sin2(\zeta_S-\phi_S)
\end{equation} 
provides the effective response functions matching condition $\chi_{{\rm eff}S}^{-1} = -\chi_{{\rm eff}I}^{-1}$.

Finally, we have to make the effective coupling factors equal to each other, which gives the following condition:
\begin{equation}\label{cond3_raw} 
  \beta_I^2 = \beta_S^2 \,.
\end{equation} 

In order to simplify the equations, it is convenient to introduce the following combined angles:
\begin{equation}
  \zeta \equiv \zeta_I - \phi_I\, ,\quad \phi \equiv \phi_I + \phi_S\, .
  \label{eq:comb-angles}
\end{equation}
In these notations, the conditions~\eqref{cond1_raw}, \eqref{cond2_raw}, and \eqref{cond3_raw} can be reexpressed and summarized as
\begin{subequations}\label{eq:quasi-optimum}
  \begin{gather}
    \zeta + \phi_I + \zeta_S = \pi \,, \label{eq:angles-sol}\\
     \Omega_{qS}^2 = \frac{\sin^2\zeta}{\sin^2(\zeta+\phi)}\,\Omega_{qI}^2 \, ,
      \label{eq:theta-S-sol}\\ 
         \Omega_S^2 = \frac{\sin\zeta\sin\phi}{\sin(\zeta+\phi)}\,\Omega_{qI}^2 \, .
    \label{eq:Omega-S-sol}
  \end{gather}
\end{subequations}
These conditions generalize those of the simpler scheme, Eqs.~\eqref{eq:theta-matching-intuitive} and~\eqref{eq:chi-matching-intuitive}, which are recovered from Eqs.~\eqref{eq:theta-S-sol} and~\eqref{eq:Omega-S-sol} as the special case $\zeta=\pi/2$ and $\phi=0$.

\subsection{Geometrical interpretation of noise cancellation using virtual rigidity\label{sec:intuitive-scheme}}

\begin{figure}[tb]
%\centering
%\def\svgwidth{0.35\textwidth}
%\input{Quadratures_v5.pdf_tex}
  \includegraphics[width=0.4\textwidth]{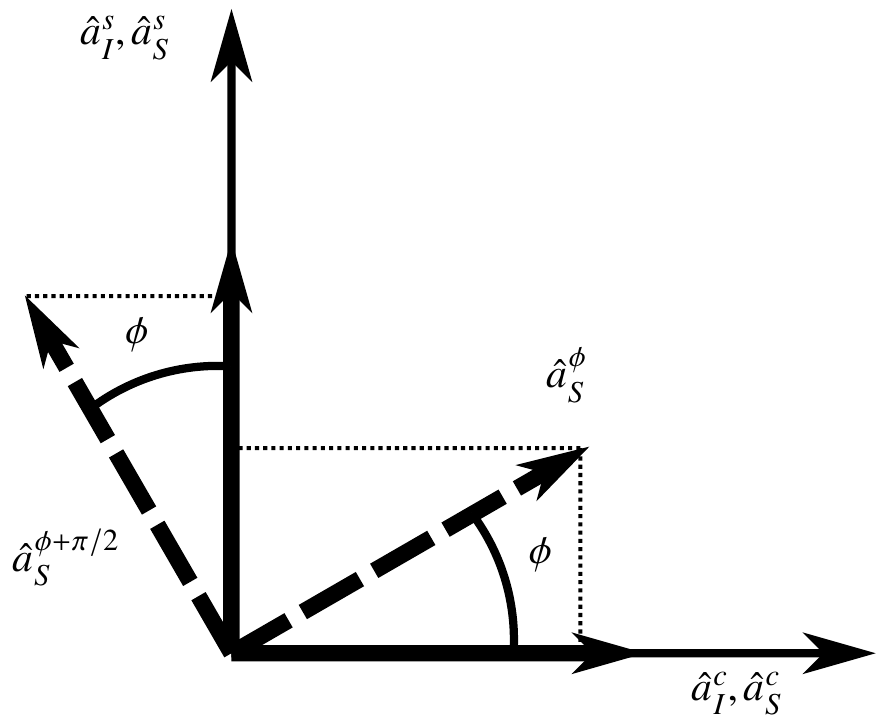}
\caption{Geometrical representation of the virtual rigidity effect. The back-action and imprecision noise quadratures for $I$ and $S$ are depicted with respect to the quadrature bases $(\hat{a}_{I,S}^{c},\hat{a}_{I,S}^{s})$, respectively. Homodyne measurement quadratures are chosen to be $\hat{b}_I^{s}$ and $\hat{b}_S^{s}$ (\ie, $\zeta_I=\zeta_S=\pi/2$) so that the corresponding imprecision noise quadratures (vertical bold arrow) coincide with the vertical axes, $\hat{a}_I^{s}$ and $\hat{a}_S^{s}$. For $I$, we make the standard choice of probe phase $\phi_I=0$ such that the orthogonal quadrature $\hat{a}_I^{c}$ acts as back action. However, for $S$, a probe phase $\phi=\phi_S$ relative to $\hat{a}_S^{c}$ is introduced.}
\label{fig:Quadratures}
\end{figure}

We now provide a geometrical interpretation of the conditions for quantum noise cancellation arrived at above. To this end, we focus on the case $\zeta=\pi/2$, which is the natural choice for purposes of broadband sensing enhancement (as will be discussed below). In view of Eqs.~\eqref{eq:comb-angles}, we may choose $\phi_I=0$ as a matter of convention, since this is without consequences for $\Omega_{qS}$ and $\Omega_S$~[Eqs.~\eqref{eq:theta-S-sol} and~\eqref{eq:Omega-S-sol}]; it follows that $\phi=\phi_S$ and, from Eq.~\eqref{eq:angles-sol}, $\zeta_S=\pi/2$. Conveniently, this implies equality between the original and primed quadrature bases $\hat{{\rm a}}_{I,S}^{c,s}$ and $\hat{{\rm a}}_{I,S}^{c,s}{}'$ [Eqs.~\eqref{a_cs_prime}].

In this scenario the back-action matching condition~\eqref{eq:theta-S-sol} can be written [cf.~Eq.~\eqref{eq:theta-matching-intuitive}]
\begin{equation}
  \Omega_{qI}^2 = \Omega_{qS}^2\cos^2 \phi = 4\Omega_S \Gamma_S \cos^2 \phi \, , 
  \label{eq:theta-matching-intuitive2}
\end{equation}
where only the component $\Gamma_S \cos^2 \phi$ of the spin back action $\hat{{\rm a}}_{S}^{\phi}$ overlapping with that of the interferometer $\hat{{\rm a}}_{I}^{c}$ contributes (Fig.~\ref{fig:Quadratures}); the projection factor is $\cos\phi$ in terms of amplitude and, thus, $\cos^2\phi$ in terms of power. 
However, the power broadening of $\gamma_S$ and associated noise are still proportional to the full back-action rate $\Gamma_S$ according to Eq.~\eqref{eq:power-broad}.

The virtual rigidity effect results in an effective spin resonance frequency $\Omega_{S\text{eff}}$ as found in Eq.~\eqref{eq:chi-eff-S}:
\begin{equation}
\Omega_{S\text{eff}}^2 = \Omega_S^2 - \frac{\Omega_{qS}^2}{2}\sin2\phi 
= \Omega_S^2 - \Omega_{qI}^2\tan\phi \, ,\label{eq:Omega_Seff}
\end{equation}
where the condition~\eqref{eq:theta-matching-intuitive2} was used to obtain the last expression. 
This shift can be understood geometrically from Fig.~\ref{fig:Quadratures} as follows: The state of $S$ is mapped with strength $\Omega_{qS}^2$ onto the output quadrature $\hat{b}_S^{\phi+\pi/2}$ whose imprecision noise $\hat{a}_S^{\phi+\pi/2}$ (orthogonal to the back action $\hat{a}_S^{\phi}$) has the projection factor $\cos\phi$ onto $\hat{a}_S^{s}$, the imprecision noise of the chosen measurement quadrature. Meanwhile the part of the back action $\hat{a}_S^{\phi}$ correlated with $\hat{a}_S^{s}$ is the projection $\sin\phi$. Combining these factors, the resulting interference is found to be equivalent to a shift in the resonance frequency squared by  $-\Omega_{qS}^2 \cos\phi\sin\phi=-(\Omega_{qS}^2/2) \sin 2\phi$, see Eq.~\eqref{eq:Omega_Seff}, where the minus sign arises from the negative effective mass, cf.~Eq.~\eqref{chis_ideal}.
Matching requires $\Omega_{S\text{eff}}=0$ as discussed above, thus, according to Eq.~\eqref{eq:Omega_Seff}, fixing
\begin{equation}
\Omega_S = \Omega_{qI}\sqrt{\tan\phi} \, ,\label{eq:Omega_S-sol}
\end{equation}
which is consistent with Eq.~\eqref{eq:Omega-S-sol} for $\zeta=\pi/2$. 

\subsection{Optimal $\phi$ for minimizing spin decay $\gamma_S$}\label{sec:minimize-spin-decay}

Continuing to focus on the choice $\zeta=\pi/2$ relevant for broadband quantum noise evasion, we now finally address to which extent the virtual rigidity effect can alleviate the limitation of the original scheme reviewed in Sec.~\ref{sec:matching_conds}. Our goal is to approach the quantum noise matching conditions while minimizing $\gamma_S$ and, hence, $\Gamma_S$; see Eq.~\eqref{eq:power-broad}. As is clear from Eq.~\eqref{eq:Omega_S-sol}, the virtual rigidity effect allows us (in principle) to work at an arbitrarily large $\Omega_S$ for a given $\Omega_{qI}$ by letting $\phi\rightarrow \pi/2$. But the projection factor $\cos^2\phi$ in Eq.~\eqref{eq:theta-matching-intuitive2} entails that  under the constraint of matching~\eqref{eq:quasi-optimum} the minimum of $\gamma_S$ occurs at a finite value of $\Omega_S$. Combining Eqs.~\eqref{eq:theta-matching-intuitive2} and~\eqref{eq:Omega_S-sol}, we find that
\begin{equation}
\Gamma_S = \frac{\Omega_{qI}}{4\sqrt{\sin\phi \cos^3\phi}} \, ,\label{eq:Gamma_S-sol}
\end{equation}
which is minimal at $\phi = \pi/6$, resulting in
\begin{equation}
\Gamma_S=\Omega_{qI}/3^{3/4}\,,\quad \Omega_S = \Omega_{qI}/3^{1/4}\, ,\label{eq:params_phi-pi-over-6}
\end{equation}
and, hence, in view of Eq.~\eqref{eq:power-broad}, the minimized spin decay $\gamma_S$.
Since this is the main bottleneck parameter, as will be clear from the full sensitivity calculation below, the parameters $\zeta=\pi/2$, $\phi=\pi/6$, and~\eqref{eq:params_phi-pi-over-6} constitute a quasi-optimal set for broadband quantum noise evasion resulting in a much less stringent requirement for the spin cooperativity $\mathcal{C}_S$ and/or the intrinsic decay rate $\gamma_{S,0}$ compared to the original proposal~\cite{18a1KhPo}.

\section{Calculation of the sensitivity}\label{sec:exact-S}

\subsection{Assumptions and approximations}\label{sec:exact-S-intro}

In the analysis above, we arrived at the conditions for perfect quantum noise cancellation while elucidating the essential physics of our scheme. However, this was done in the idealized limit of dissipationless mechanical and spin degrees of freedom. While the mechanical losses in modern gravitational wave detectors are very small and can safely be neglected in our analysis, the imperfections brought about by finite bandwidth $\gamma_S$ of the spin system as well as by the optical losses must be accounted for in order to assess the feasibility of the scheme. 

A finite $\gamma_S$ will impact performance in two ways. First, it will introduce a non-zero imaginary part to the spin system susceptibility, 
\begin{equation}\label{chi_S_loss} 
  \chi_{S}^{-1} =-(\Omega_S^2 - \Omega^2 - 2i\Omega\gamma_S) \,,
\end{equation} 
which cannot be countered by the virtual rigidity effect and, hence, will render perfect broadband matching of the effective susceptibilities $\chi_{\text{eff}I}^{-1}$ and $\chi_{\text{eff}S}^{-1}$ impossible. Second, on account of the dissipation-fluctuation theorem, spin noise uncorrelated with the meter noise will degrade the sensitivity.

The optical losses can in a natural way be divided into three parts: the losses in the input paths of the interferometer and the spin system, the light absorption inside these two subsystems (\eg, the intracavity losses), and the losses in their output paths. The input losses include, in particular, the imperfections of the squeezer, whereas the output losses include, in particular, the finite quantum efficiency of the detectors. 
In order to simplify the equations, we neglect the intracavity losses. In the contemporary GW interferometers, the influence of these losses is small in comparison with the input and the output counterparts. Concerning the spin system, the virtual rigidity approach requires only the modest quantum cooperativity $C_S\sim10$, which allows the use of a small-finesse cavity or even a cavityless path-through topology, which also makes the internal optical losses much smaller than those at the input and the output. 

In our analysis here and below, we still assume the bad-cavity approximation~\eqref{bad_cavity}. This approach is justified by the following reasoning. The spectral shape of the quantum noise of interferometers depends on the two characteristic frequencies $\Omega_q$ and $\gamma$. Below $\Omega_q$, the radiation pressure noise dominates over the imprecision shot noise (if no back-action cancellation techniques are used). Above $\gamma$, the normalized noise increases due to the signal cutting by the interferometer bandwidth. 
In all contemporary and planned GWDs, $\Omega_q\ll\gamma$. If frequency-independent squeezing is used, then the effective $\Omega_q$ scales up as $e^{2r}$. However, in frequency-{\it dependent} squeezing schemes like the one considered here, both the back-action noise and the imprecision shot noise are suppressed, leaving $\Omega_q$ unchanged. Therefore, all non-trivial behavior of the back-action cancellation schemes is concentrated in the frequency band $\Omega\sim\Omega_q\ll\gamma$. In particular, it is easy to show that, at higher frequencies, the scheme which we consider here provides only the trivial (but desirable) frequency-independent suppression of the sum quantum noise, which consists only of the imprecision shot noise in this frequency band, by a frequency-independent factor defined by the squeezing rate and the optical losses. 

\subsection{Effects of dissipation and losses}\label{sec:total-noise}

With account of optical losses and spin dissipation, the input-output relations for the interferometer and the spin system take the following form:
\begin{subequations}
  \begin{multline}
    \hat{b}_I^{\zeta_I} = \sqrt{\eta_{oI}}\biggl[
        \hat{a}_I^{\zeta_I}
        + \chi_I\biggl(
              \frac{\Omega_{qI}}{\sqrt{\hbar m}}\,F_s + \Omega_{qI}^2\hat{a}_{I}^{\phi_I}
            \biggr)\sin(\zeta_I-\phi_I)
      \biggr] \\
      + \sqrt{1-\eta_{oI}}\,\hat{z}_I^{\zeta_I} \,, 
  \end{multline}
  \begin{multline}
    \hat{b}_S^{\zeta_S} = \sqrt{\eta_{oS}}\bigl[
        \hat{a}_S^{\zeta_S}
        + \chi_S\bigl(\Omega_{qS}\hat{f}_T + \Omega_{qS}^2\hat{a}_S^{\phi_S}\bigr)
            \sin(\zeta_S-\phi_S)
      \bigr] \\
      + \sqrt{1-\eta_{oS}}\,\hat{z}_S^{\zeta_S} \,, 
  \end{multline}
\end{subequations}
compare with Eq.~\eqref{b_zeta}. Here $\hat{z}_{I,S}$ are the vacuum fields associated with the output optical losses of the interferometer and the spin system channels and  $\hat{f}_T$ is the normalized thermal force of the spin system with the spectral density defined by the fluctuation-dissipation theorem:
\begin{equation}
  \sigma_T \ge 2\gamma_S\Omega \,.\label{eq:sigma-T}
\end{equation} 
We assume that this noise is ground-state noise, \ie, with equality in the above equation. Accounting for the input optical losses, the spectral densities and correlations~\eqref{eq:entanglement-corr} of the incident optical fields generalize to
\begin{subequations}\label{eq:entanglement-corr-general-main-text}
  \begin{gather}
    S_{a_I^c} = S_{a_I^s} = \eta_{iI}\sh^2r + \nf{1}{2} \,, \\
    S_{a_S^c} = S_{a_S^s} = \eta_{iS}\sh^2r + \nf{1}{2} \,, \\
    S_{a_I^ca_S^c} = -S_{a_I^sa_S^s} = \frac{1}{2}\sqrt{\eta_{iI}\eta_{iS}}\sh2r \,.
  \end{gather}
\end{subequations}

Similar to our treatment in Sec.~\ref{sec:noise-cancel-intuitive}, we introduce the normalized noise forces
\begin{subequations}\label{f_IS_loss} 
  \begin{gather}
    \hat{f}_I 
      = \frac{
            \chi_{\text{eff}I}^{-1}\hat{a}_I^s{}'
            + \chi_I^{-1}\epsilon_{oI}\hat{z}_I^{\zeta_I} 
          }{\beta_I}\, 
        + \beta_I\hat{a}_I^c{}'  \,, \\
    \hat{f}_S 
      = \frac{
            \chi_{\text{eff}S}^{-1}\hat{a}_S^s{}'
            + \chi_S^{-1}\epsilon_{oS}\hat{z}_S^{\zeta_S} 
          }{\beta_S}\, 
        + \beta_S\hat{a}_S^c{}' + \hat{f}_T \, ,
  \end{gather}
\end{subequations}
where 
\begin{equation}
  \epsilon_{oI,S} = \sqrt{\frac{1-\eta_{oI,S}}{\eta_{oI,S}}} \, ;
\end{equation} 
compare with Eqs.~\eqref{f_IS_rotated}. 
The spectral densities of $\hat{f}_I$ and $\hat{f}_S$ and their cross-spectral density are equal to
\begin{subequations}\label{sigma_general}
  \begin{gather}
    \sigma_I = \frac{\eta_{iI}K_I}{2}(\ch2r + \varkappa_I) \,,\\
    \sigma_S = \frac{\eta_{iS}K_S}{2}(\ch2r + \varkappa_S) \,, \\
    \sigma_{IS} = \frac{1}{2}\sqrt{\eta_{iI}\eta_{iS}}\,K_{IS}\sh2r \,,
  \end{gather}
\end{subequations}
where we have introduced the factors
\begin{subequations}
  \begin{gather}
    \varkappa_I = \epsilon_{iI}^2 + \frac{k_I\epsilon_{oI}^2}{K_I\eta_{iI}} \,,\\
    \varkappa_S = \epsilon_{iS}^2 + \frac{k_S\epsilon_{oS}^2 + 2\sigma_T}{K_S\eta_{iS}}
      \,,\label{eq:varkappaS}
  \end{gather}
\end{subequations}
describing the total imperfections in each of the channels, and the coefficients
\begin{subequations}\label{Ks_def} 
  \begin{gather}
    K_{I,S} = \frac{|\chi_{\text{eff}I,S}^{-1}|^2}{\beta_{I,S}^2} + \beta_{I,S}^2 \,,\\
    K_{IS} = K_{IS}^c\cos(\zeta_I+\zeta_S) + K_{IS}^s\sin(\zeta_I+\zeta_S) \,, \\
    k_{I,S} = \frac{|\chi_{I,S}^{-1}|^2}{\beta_{I,S}^2} \, ,
  \end{gather}
\end{subequations}
with
\begin{equation}
  K_{IS}^c = \frac{\chi_{\text{eff}I}^{-1}\chi_{\text{eff}S}^{-1\ast}}{\beta_I\beta_S}
    - \beta_I\beta_S \,, \quad 
  K_{IS}^s = \frac{\beta_S}{\beta_I}\chi_{\text{eff}I}^{-1}
    + \frac{\beta_I}{\beta_S}\chi_{\text{eff}S}^{-1*}  \,.
\end{equation} 

The optimally combined sum noise spectral density of the two meters is given by
\begin{multline}
  \frac{S}{\hbar m} = \sigma_I - \frac{|\sigma_{IS}|^2}{\sigma_S} 
  = \frac{\eta_{iI}}{2K_S(\ch2r + \varkappa_S)}\bigl\{
        |K_{\rm res}|^2\ch^22r \\
        + |K_{IS}|^2 
        + K_IK_S[(\varkappa_I+\varkappa_S)\ch2r + \varkappa_I\varkappa_S]
      \bigr\} , \label{eq:S-general}
\end{multline}
where
\begin{equation}
  |K_{\rm res}|^2 = K_IK_S - |K_{IS}|^2 \,.
\end{equation} 
Note that the structure of Eq.~\eqref{eq:S-general} suggests that an optimal value of squeezing providing the minimum of $S$ should exist.  This optimization is done in the following subsection.

\subsection{Squeezing optimization}\label{sec:opt-r}

Here we derive the optimal value of the squeezing parameter $r$ based on the general expressions given in Sec.~\ref{sec:total-noise}. To be precise, we derive the value of $r$ that minimizes the sensitivity $S(\Omega)$~\eqref{eq:S-general} at a given Fourier component $\Omega$. 
Non-zero squeezing (and, hence, introducing the spin system) can improve the sensitivity only if the extraneous noise in the spin system uncorrelated with the interferometer is sufficiently small; from Eq.~\eqref{eq:S-general}, we find the following condition on $\varkappa_S$~\eqref{eq:varkappaS}:
\begin{equation}
\varkappa_S < 1 - 2\frac{|K_{\text{res}}|^2}{K_I K_S}\, .\label{eq:sq-cond}
\end{equation}
Recall that these quantities depend on the Fourier frequency $\Omega$.
Provided that the condition~\eqref{eq:sq-cond} is fulfilled, the squeeze parameter $r_{\text{opt}}(\Omega)$ that minimizes the sensitivity $S(\Omega)$ is specified by 
\begin{equation}
  \cosh 2r_{\text{opt}}(\Omega) 
  = \frac{|K_{IS}|}{|K_{\text{res}}|}\sqrt{1-\varkappa_S^2} - \varkappa_S \, .
  \label{eq:r-opt}
\end{equation}
Note that Eqs.~\eqref{eq:sq-cond} and~\eqref{eq:r-opt} are both independent of the extraneous noise in the \emph{interferometer} due to optical losses, $\eta_{iI},\eta_{oI}<1$.
If the source of entangled light is broadband, as assumed throughout this work, it is characterized by a squeeze factor $r$ which is independent of the Fourier frequency; accordingly, the chosen $r$ will not be optimal for all Fourier frequencies $\Omega$. With this in mind, we may nonetheless evaluate $S(\Omega)$~\eqref{eq:S-general} at $r=r_{\text{opt}}(\Omega)$ to achieve an expression for the optimized sensitivity at each Fourier frequency:
\begin{multline}
%  \frac{S_{\text{opt}}}{\hbar m} \equiv
   \left.\frac{S}{\hbar m}\right|_{r=r_{\text{opt}}}\hspace*{-0.7em}
  =\frac{\eta_{iI}}{2K_S}\left[2 |K_{\text{res}}|^2 \cosh 2r_{\text{opt}} 
    + K_IK_S(\varkappa_I+\varkappa_S)\right] ,
\end{multline}
with $\cosh 2r_{\text{opt}}$ given by Eq.~\eqref{eq:r-opt}, suppressing the $\Omega$ dependence of the involved quantities for brevity.

\subsection{Quasi-optimal noise cancellation in the large-squeezing limit}

We now optimize the full sensitivity~\eqref{eq:S-general} in the large-squeezing limit $e^{2r}\gg1$. While the parameters which provide this optimum are not strictly the best ones for finite squeezing and in the presence of optical and spin losses and noise, it is still a reasonable working point when these detrimental effects are relatively small. 

In the limit of $r\to\infty$ Eq.~\eqref{eq:S-general} simplifies to 
\begin{equation}\label{S_big_r} 
  \frac{S}{\hbar m} 
  = \frac{\eta_{iI}e^{2r}}{4}\,\frac{|K_{\rm res}|^2}{K_S} \, .
\end{equation} 
In order to minimize this expression, the cross-correlation term $|K_{IS}|$ has to be maximized. In principle, its rigorous maximization in $\zeta_I+\zeta_S$ is possible, but its maximum corresponds to frequency-dependent homodyne angles. We assume instead that the condition \eqref{cond1_raw} is fulfilled. In this case,
\begin{equation}
  |K_{\rm res}|^2 = \left|
      \frac{\beta_S}{\beta_I}\chi_{\text{eff}I}^{-1}
      + \frac{\beta_I}{\beta_S}\chi_{\text{eff}S}^{-1}
    \right|^2 .
\end{equation} 
It is easy to see that, in the ideal lossless case of \eqref{chis_ideal}, the  conditions~\eqref{cond2_raw} and \eqref{cond3_raw} make $K_{\rm res}$ equal to zero.

Unfortunately, the imaginary part of the realistic spin system susceptibility \eqref{chi_S_loss} does not allow us to cancel $K_{\rm res}$ completely. Taking it into account and still assuming the conditions~\eqref{cond1_raw}, \eqref{cond2_raw}, and \eqref{cond3_raw}, we obtain that
\begin{subequations}
  \begin{gather}
    K_S = K_I + \frac{4\Omega^2\gamma_S^2}{\beta_I^2} \,,\\
    |K_{IS}|^2 = K_I^2 + \frac{4\chi_{\text{eff}I}^{-2}\Omega^2\gamma_S^2}{\beta_I^4}\,,\\
    |K_{\rm res}|^2 = 4\Omega^2\gamma_S^2 \,. 
  \end{gather}
\end{subequations}
Note that $\chi_{\text{eff}I}$ depends only on the angle $\zeta$, as is clear from Eqs.~\eqref{eq:chi-eff} and~\eqref{eq:comb-angles}. We thus remark that the optimized noise spectral density~\eqref{eq:S-general} depends on the angle $\phi$ only via $\chi_S^{-1}$, as this dictates our choice of $\Omega_S$ according to Eq.~\eqref{eq:Omega-S-sol}. The only term in Eq.~\eqref{eq:S-general} that contains $\chi_S^{-1}$ represents the vacuum noise due to optical losses at the output of the spin system and is minimal at the peak of $\chi_S(\Omega)$ ($\Omega\sim\Omega_S$ provided that $\Omega_S \gg \gamma_S$). Hence, in the absence of such losses, $\eta_{oS}\rightarrow 1\Rightarrow \epsilon_{oS}\rightarrow 0$, the sensitivity is independent of both $\phi$ and $\Omega_S$. Crucially, this allows a significant extent of freedom in choosing the (bare) spin resonance $\Omega_S$ since the appropriate effective (virtual) resonance can achieved by the proper choice of $\phi$; this is a great advantage for purposes of practical implementation and is a central result of this work.

\section{Discussion of the sensitivity}\label{sec:plots}

We now explore this central idea in a case of particular interest $\zeta=\pi/2$, so that $\chi_{\text{eff}I}=\chi_{I}$, which means that the standard phase configuration of orthogonal probe and homodyne quadratures is used for the main interferometer; this is exactly the scenario discussed in Sec.~\ref{sec:intuitive-scheme}. As for an interferometer without a spin system, this yields the most flat shape of the sum quantum noise spectral density, as well as the best performance in the shot-noise-dominated high-frequency band. Moreover, in all plots we choose $\phi=\pi/6$ in order to minimize $\gamma_S$ for fixed $\mathcal{C}_S$ according to the discussion in Sec.~\ref{sec:minimize-spin-decay}; hence, the quasi-optimal $\Omega_S$ and $\Gamma_S$ are given as functions of $\Omega_{qI}$ alone; see Eqs.~\eqref{eq:params_phi-pi-over-6}.

\begin{figure}[tb]
\begin{center}
   \includegraphics[width=\columnwidth]{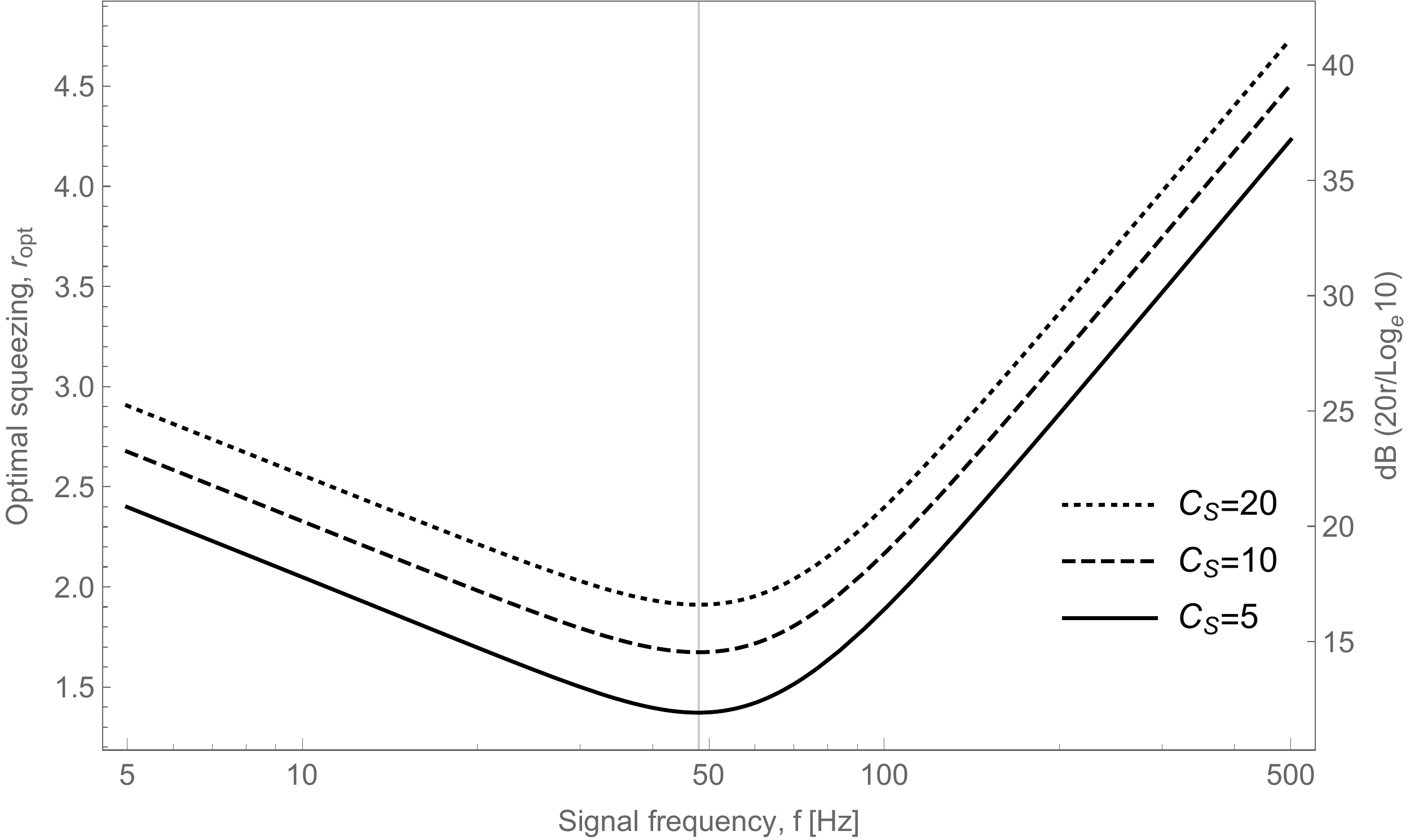} 
\caption{Optimal squeezing as function of Fourier frequency for different values of the spin cooperativity $\mathcal{C}_S$. The bare spin resonance $\Omega_S/(2\pi)$ is indicated by the vertical line and coincides to good approximation with the minimum of $r_{\text{opt}}(\Omega)$.} %\ez{The decibel scale is defined as $r_{\text{dB}}=20r/\ln 10$.}
\label{fig:r-opt}
\end{center} 
\end{figure}

Prior to plotting the sensitivity resulting from these parameters, let us discuss the choice of squeezing $r$ in the presence of imperfections. The optimal degree of two-mode squeezing $r_{\text{opt}}$ depends on the signal Fourier frequency $\Omega$; see Sec.~\ref{sec:opt-r}. This function is plotted in Fig.~\ref{fig:r-opt} for different spin cooperativities $\mathcal{C}_S$ and exhibits a minimum near $\Omega \sim \Omega_S$ for the present parameters, as will be discussed below in this section. 

Taking into account that in GWDs the main interest is in broadband sensing enhancement, we will here choose the squeezing $r=r_{\text{min}}$ that maximizes the minimal sensitivity gain within the signal bandwidth. This will tend to flatten the gain curve and will in some sense represent a conservative assessment of the scheme in that the chosen squeezing is optimal only for a single point on the gain curve.

The degree of two-mode squeezing $r$, defined in Eqs.~\eqref{eq:entanglement-corr} and shown in Fig.~\ref{fig:r-opt}, corresponds to the pure state generated by the optical parametric oscillator and not to what is actually observed given the losses and imperfections which are taken into account separately. Therefore, it essentially is limited only by the pumping power. More specifically, the expression for the degree of squeezing in a pure state as a function of the ratio of the pump power to the threshold pump power, $P/P_\text{th}$, is $\exp 2r = (1+\sqrt{P/P_\text{th}})^2/(1-\sqrt{P/P_\text{th}})^2$~\cite{Polzik1992}. 
It follows that the degree of squeezing of 15 dB is achieved at half threshold power, and 20 dB can be achieved at 67\% of threshold power, which is well within the experimental reach.

\begin{figure}[tb]
\begin{center}
   \includegraphics[width=\columnwidth]{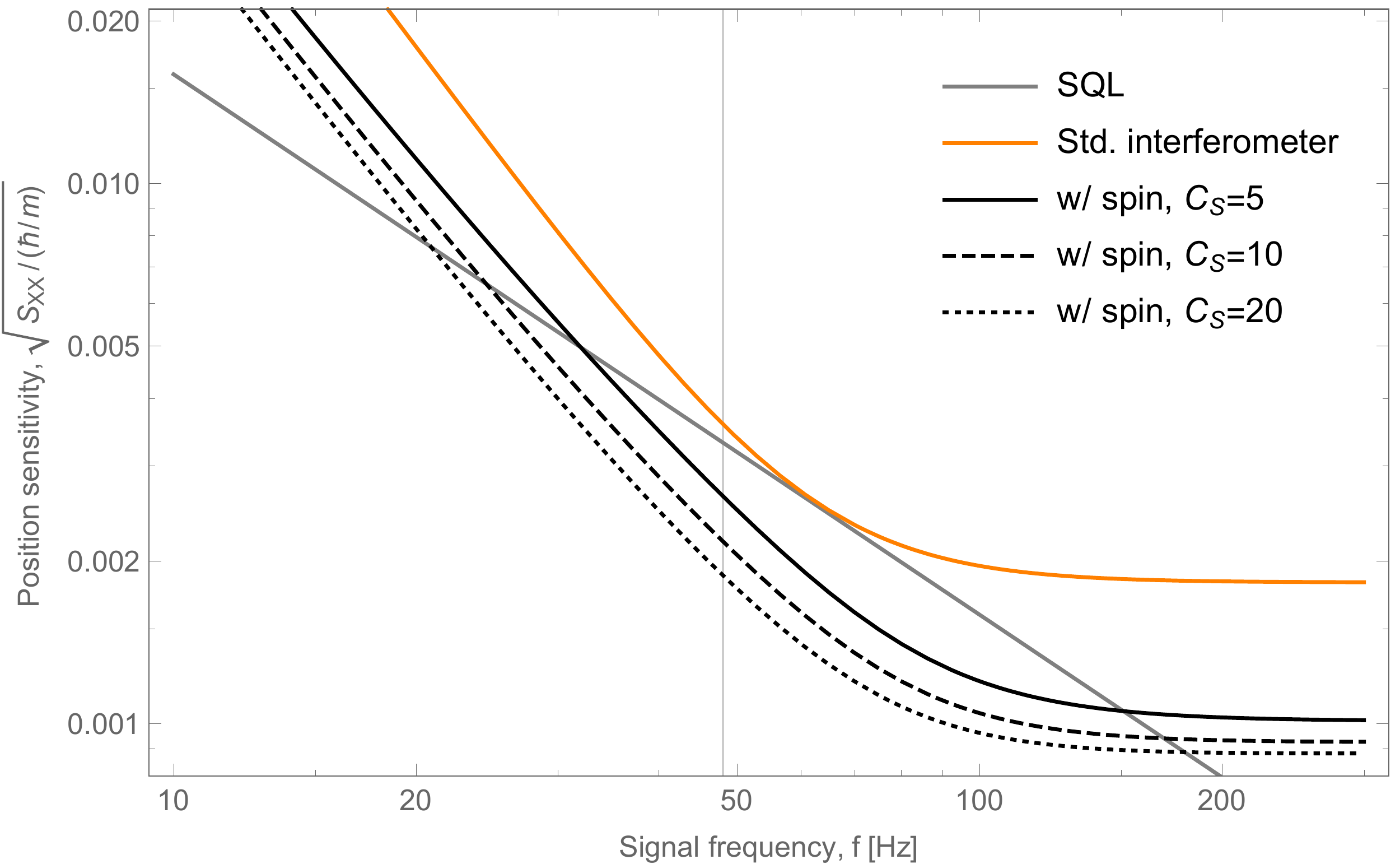} 
\caption{Normalized position sensitivity $\sqrt{|\chi_I(\Omega)|^2 S}$ for spin-GWD scheme benchmarked against standard interferometer. Also shown is the standard quantum limit (SQL).}
 \label{fig:sens-FF}
\end{center}
\end{figure}

In Fig.~\ref{fig:sens-FF}, the normalized quantum noise spectral density is plotted for the parameters discussed above and for three realistic values of $\mathcal{C}_S$. For comparison, the spectral density $S_\text{std}$ of the ``standard'' interferometer, which corresponds to $r\rightarrow 0$ and $\zeta=\pi/2$, is also shown in this plot. It can be seen that a broadband sensing gain is possible relative to a standard interferometer, surpassing the SQL over a broad band, as also demonstrated in the original proposal~\cite{18a1KhPo}. 

\begin{figure}[thb]
\begin{center}
   \includegraphics[width=\columnwidth]{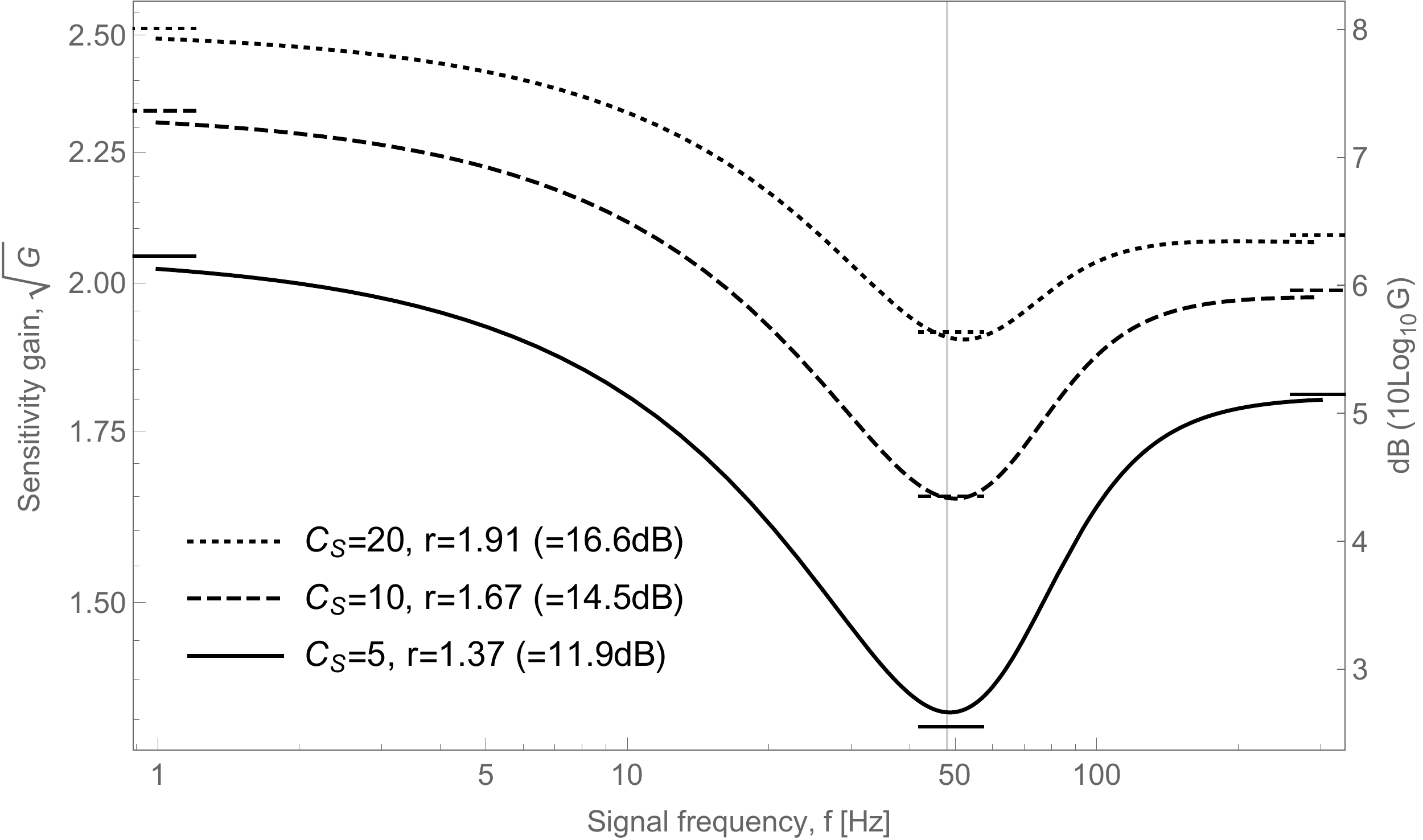} 
\caption{Sensitivity gain relative to standard interferometer $\sqrt{G}\equiv\sqrt{S_{\text{std}}/S}$. The vertical line indicates $\Omega_S/(2\pi)$. Gain values predicted by simplified expressions for low~\eqref{eq:Gapp-LF} and high~\eqref{eq:Gapp-HF} frequencies are indicated by horizontal line segments. We also indicate the gain at $\Omega=\Omega_S$, near the minimum, according to the approximate formula Eq.~\eqref{Gapp}.}
\label{fig:sens-gain}
\end{center} 
\end{figure}

To further illustrate the sensitivity gain relative to the standard interferometer, we introduce the gain factor
\begin{multline}
  G = \frac{S_{\rm std}}{S} =\\
   \frac{K_IK_S(1+\varkappa_I)(\ch2r + \varkappa_S)}{
        |K_{\rm res}|^2\ch^22r + |K_{IS}|^2 
          + K_IK_S[(\varkappa_I+\varkappa_S)\ch2r + \varkappa_I\varkappa_S]
      } \,. \label{G}
\end{multline} 
It is plotted in Fig.~\ref{fig:sens-gain} for the same values of the parameters as in Fig.~\ref{fig:sens-FF}. Note that since the event rate of GWDs scales as the volume within which the GWD is sensitive, it is proportional to $G^{3/2}$. 

Equation \eqref{G} is quite cumbersome. However, for the reasonable values of the parameters which we use for our estimates, it can be significantly simplified, providing insight into the comparative influence of the optical losses and the spin system bandwidth on the sensitivity. Focusing on the case $\zeta=\pi/2$ and making the reasonable assumptions that $\varkappa_{I,S}\ll\ch2r$, $\gamma_S \ll \Omega_{qI}$, and $\epsilon_{iS}^2 \ll 1$, Eq.~\eqref{G} can be approximated as
\begin{equation}
%  G \approx G_{\rm approx} = \frac{(1+\varkappa_I)\cosh2r}{
%      \left(1+\dfrac{2\gamma_S \Omega \ch2r}{K_I}\right)^2 
%      + (\varkappa_I+\varkappa_{S\,{\rm opt}})\ch2r} 
  G \approx G_{\rm approx} = \frac{1+\varkappa_I}{
      \left(\dfrac{1}{\sqrt{\ch2r}}+\dfrac{2\gamma_S \Omega \sqrt{\ch2r}}{K_I}\right)^2 
      + \varkappa_I+\varkappa_{S\,{\rm opt}}} 
      \,, \label{Gapp}
\end{equation} 
where 
\begin{equation}
  \varkappa_{S\,{\rm opt}} = \epsilon_{iS}^2 + \frac{k_S\epsilon_{oS}^2}{K_S\eta_{iS}}
\end{equation} 
is the part of $\varkappa_S$ imposed by the optical losses (note that it still depends on $\gamma_S$ through the factor $k_S$, but this dependence is relatively weak). 

Equation~\eqref{Gapp} succinctly expresses the impact of spin decay $\gamma_S$, optical losses $\varkappa_I$ and $\varkappa_{S\text{opt}}$, and finite squeezing $r$ on the performance of our scheme. It shows that, at the price of introducing the additional optical losses $\varkappa_{S\text{opt}}$, the essential quantum noise contribution (\ie, in the absence of losses) is reduced by the factor given by the parenthesis squared in the denominator of Eq.~\eqref{Gapp}. The squeeze parameter $r_{\text{opt}}(\Omega)$ that maximizes $G_{\rm approx}$ at a given Fourier frequency $\Omega$ is simply given by $\cosh 2r_{\text{opt}}(\Omega)=K_I/\sigma_T$ [using Eq.~\eqref{eq:sigma-T} with equality]; this is consistent with the general result~\eqref{eq:r-opt} within the regime of validity of $G\approx G_{\rm approx}$ insofar as $\varkappa_S^2 \ll 1$.

Equation~\eqref{Gapp} clearly shows also that the characteristic dip in the frequency dependence of $G$ (see Fig.~\ref{fig:sens-gain}) is created by the spin system damping and corresponds, to good approximation, to the maximum of the ratio $\Omega/K_I$, which occurs at the frequency 
\begin{equation}
  \Omega_{\rm min} = \Omega_{qI}/3^{1/4} \, .
\end{equation} 
It can be seen from Eq.~\eqref{eq:params_phi-pi-over-6} that $\Omega_{\rm min}$ coincides with the bare spin resonance $\Omega_{\rm min}=\Omega_S$ when $\phi=\pi/6$, the choice that minimizes $\gamma_S$. For our values of the parameters, it evaluates to $\Omega_{\rm min}\approx2\pi\times 48\,{\rm Hz}$. Estimates show that around this frequency, the value of $G$ is limited mostly by the spin system damping. 
Due to the increase of the spin system damping influence around $\Omega_{\rm min}$, a smaller degree of squeezing (that is, less entangled optical fields in the interferometer and the spin system channels) becomes optimal in this frequency band, creating the aforementioned dip in Fig.~\ref{fig:r-opt}.

At the same time, for low and high signal frequencies, $K_I\gg2\gamma_S\Omega\ch2r$ and 
\begin{equation}\label{Gapp2}
  G \approx \frac{1+\varkappa_I}
    {1/\ch2r + \varkappa_I+\varkappa_{S\,{\rm opt}}} \, ;
\end{equation} 
that is, the sensitivity gain is primarily defined by the optical losses. Keeping only the lowest-order terms in $\epsilon_{iI,S}^2$ and $\epsilon_{oI,S}^2$ in the numerator and denominator of Eq.~\eqref{Gapp2}, we find for low (back-action-dominated) frequencies $\Omega\ll\Omega_{qI}$
\begin{equation}\label{eq:Gapp-LF}
  G \approx \frac{1+\epsilon_{iI}^2}{
      1/\ch2r + \epsilon_{iI}^2 + \epsilon_{iS}^2 + \epsilon_{oS}^2\tan^2\phi
    } \, ,
\end{equation} 
whereas for high (imprecision-noise-dominated) frequencies $\Omega\gg\Omega_{qI}$
\begin{equation}\label{eq:Gapp-HF}
  G \approx \frac{1+\epsilon_{iI}^2+\epsilon_{oI}^2}{
      1/\ch2r + \epsilon_{iI}^2 + \epsilon_{oI}^2 
      + \epsilon_{iS}^2 + \epsilon_{oS}^2
    } \, .
\end{equation} 
The approximate Eqs.~\eqref{eq:Gapp-LF} and~\eqref{eq:Gapp-HF} are compared to the exact sensitivity gain in Fig.~\ref{fig:sens-gain}.

\section{Conclusion and outlook}\label{sec:conclusion}

We have discussed a scheme that promises to push gravitational wave detectors into a new realm of broadband sub-SQL sensitivity by means of a flexible and unintrusive extension of existing interferometer topologies by a negative effective mass spin oscillator. The physical resources required by this approach are relatively low cost compared to other candidate techniques for quantum noise evasion proposed for future GWDs.

We have shown that the addition of the virtual rigidity technique to the scheme first proposed in Ref.~\cite{18a1KhPo} allows one to achieve a tangible sensitivity gain even for modest values of the atomic cooperativity, $\mathcal{C}_S \sim 10$, which dramatically facilitates practical implementation of this scheme.

The value of the resulting sensitivity crucially depends on two factors: optical losses in the scheme and the dissipation rate in the atomic spin system. For the reasonably optimistic values of these parameters used for our estimates, the sensitivity gain (in comparison with an ``ordinary'' interferometer) could reach 6--7\,dB.

While our modeling and assessment of the scheme is reasonably detailed, it is beyond the scope of the present work to exhaustively account for all realistic imperfections of gravitational wave interferometers. A first step towards the practical implementation of the scheme would be a proof-of-principle demonstration of the spin subsystem subject to a two-mode-squeezed input field. The modular nature of the setup allows such separate characterization and optimization of the spin subsystem. Given a spin system fine-tuned in this manner, work towards integration with an actual GWD could commence.

\begin{acknowledgments}
The authors would like to thank S.\,L.\ Danilishin for reading the manuscript. 
This research was supported by the Templeton Foundation, European Research Council Advanced Grant project Quantum-N and Innovation Fund Denmark (Eureka project Q-GWD). E.\,Z.\ acknowledges funding from the Carlsberg Foundation. 
\end{acknowledgments}

\end{document}